# Biaxial strain in atomically thin transition metal dichalcogenides

Riccardo Frisenda*[1], Robert Schmidt[2], Steffen Michaelis de Vasconcellos[2], Rudolf Bratschitsch[2], David Perez de Lara[1] and Andres Castellanos-Gomez*[3]

[1] Instituto Madrileño de Estudios Avanzados en Nanociencia (IMDEA-Nanociencia), Campus de Cantoblanco, E-28049 Madrid, Spain.

[2] Institute of Physics and Center for Nanotechnology, University of Münster, 48149 Münster, Germany.

[3] Instituto de Ciencia de Materiales de Madrid (ICMM-CSIC), Campus de Cantoblanco, E-28049 Madrid, Spain.

*E-mail: riccardo.frisenda@imdea.org, andres.castellanos@csic.es.

**ABSTRACT:** Strain engineering in single-layer semiconducting transition metal dichalcogenides aims to tune their bandgap energy and to modify their optoelectronic properties by the application of external strain. In this paper we study transition metal dichalcogenides monolayers deposited on polymeric substrates under the application of biaxial strain, both tensile and compressive. We can control the amount of biaxial strain applied by letting the substrate thermally expand or compress by changing the substrate temperature. After modelling the substrate-dependent strain transfer process with a finite elements simulation, we performed micro-differential spectroscopy of four transition metal dichalcogenides monolayers ($MoS_2$, $MoSe_2$, $WS_2$, $WSe_2$) under the application of biaxial strain and measured their optical properties. For tensile strain we observe a redshift of the bandgap that reaches a value as large as 94 meV/% in the case of single-layer $WS_2$ deposited on polypropylene. The observed bandgap shifts as a function of substrate extension/compression follow the order $WS_2 > WSe_2 > MoS_2 > MoSe_2$.

**Introduction**

Two-dimensional (2D) semiconductors have exciting properties that make them interesting for the construction of novel optoelectronic devices. Among the various advantages, compared to conventional bulk semiconductors, one finds high responsivity [1, 2], reduced dimensions and a large susceptibility to different external stimuli such as strain or electric fields [3, 4]. The tuning of the bandgap in bulk semiconductors through the application of strain is a practice applied in the semiconductor industry since decades. In bulk semiconductors the amount of strain applied is a fixed quantity, which is determined during the fabrication of the device and cannot be tuned as a function of time. Moreover, strain is a discrete quantity since it depends on the lattice constants of the materials used and only a limited amount of different values are available. On the other hand 2D semiconductors offers new ways of applying strain directly, both uniaxially and biaxially [5-7], which can be tuned in time in a continuous way and it is material-independent [5]. Theory also predicts that 2D materials can stand large values of strain up to 10% without breaking [8, 9], compared to approximately 1% in bulk semiconductors.

The family of transition metal dicalchogenides, whose general formula is $MX_2$ where M is a transition metal atom and X is a chalcogen atom, contains many members that can be studied in the atomic thickness limit. Among them, molybdenum and tungsten based ones ($MoS_2$, $MoSe_2$, $WS_2$, $WSe_2$) are probably the most studied [1]. These four materials are semiconductors with a bandgap in the visible, ranging from 1 to 2 eV, which is indirect in bulk and becomes direct in the monolayer [10]. The absorption properties in the visible are dominated by excitons which, due to their large binding energy, are observable even at room temperature [11]. Experiments on $MoS_2$ single-layer and few-layer flakes have already demonstrated that the optical band gap is tunable by -50 meV/% for uniaxial strain and -100 meV/% for biaxial strain [7, 12-14]. Uniaxial strain in single-layer $WS_2$ gave lower values of -20 meV/%, while similar experiments in single-layer $MoSe_2$ and $WSe_2$ gave respectively -27 meV/% and -54 meV/%. Biaxial strain in single-layer flakes of these materials was investigated only recently [15].

In this article we investigate the optical properties of Mo and W based TMDCs monolayers under the application of biaxial strain. To apply the strain we exploit the thermal expansion or compression of the substrate carrying the material. We explore three substrates characterized by different thermal expansion coefficients and Young's moduli, namely polydimethilsiloxane (PDMS), polypropylene (PP) and glass. We monitor the change in the bandgap of the TMDCs monolayers with differential reflectance spectroscopy, which in these cases gives a signal that is directly proportional to the absorption of the thin material. Studying monolayers deposited on three different substrates aid us in decoupling the effect of temperature and of the Young's moduli mismatch. We find that the amount of strain transferred from the substrate to the 2D flake is strongly dependent on the Young's modulus of the substrate and that the bandgap tunability of $MoS_2$ after discarding thermal effects is -22.1 meV/% when using PP (which has a Young's modulus 100 times lower than $MoS_2$) while it is negligible in the case of PDMS (whose Young's modulus is 100000 lower than $MoS_2$). Using PP we find that for increasing tensile strain a redshift of the optical band gap of these 2D TMDCs is observed, that reaches, in the case of $WS_2$, the value of 94 meV/%. The observed bandgap shifts as a function of substrate extension/compression follow the order $MoSe_2 < MoS_2 < WSe_2 < WS_2$, *i.e.* with $WS_2$ providing the largest bandgap tunability and $MoSe_2$ the lowest.

**Results and discussions**

To monitor the optical properties of single-layer TMDCs we employ a micro-reflectance setup, schematically depicted in Figure 1 and described in detail elsewhere [16]. Briefly, the setup is based on a Motic BA310 metallurgical microscope with a modified trinocular port and a CCD spectrometer. A white halogen lamp is used to illuminate a circular area of the sample of approximately 60 µm of diameter and the light reflected from an area of 2 µm of diameter is collected by a fiber optic and then sent to the spectrometer. Figures 1b shows a microscope photograph of the analyzer spot size and Figure 1c the intensity profiles along two directions. The sample under study can be moved with a motorized x-y stage with micrometric precision which allows collecting spatially resolved reflectance maps. A Matlab routine controls the motorized x-y stage and a power source which acts as a trigger for the spectrometer.

To isolate single-layer flakes of TMDCs we mechanically exfoliate the chosen material initially on Nitto tape and then we transfer the flakes from the tape to PDMS. The flakes that we study with the micro-reflectance setup typically have lateral dimensions much larger than the analyzed area. Figure 1d shows a microscope photograph of a $MoS_2$ flake, recorded in transmission illumination mode, onto PDMS. The flake presents regions with different colors which correspond to different values of thickness that can be identified by quantitative optical analysis and Raman spectroscopy [17]. A single-layer region, highlighted by dashed lines for clarity, is visible whose lateral dimensions are on the order of tens of micrometers, much larger than the 2 µm analyzed area. To study the flake on different substrates than PDMS we use an all-dry deterministic transfer technique which allows the transfer of a flake from PDMS to a different substrate without damaging the sample and with low residues [18]. The bottom panel of Figure 1d shows the MoS$_2$ flake transferred onto a PP substrate. Notice that because of the transfer, the flake appears mirrored.

Figure 2a shows a microscope photograph of a few WSe$_2$ flakes exfoliated onto PDMS. Similarly to the previous case of MoS$_2$, the thinner regions of the flakes are characterized by a lighter color in respect to thicker regions. To record a differential reflectance spectrum of WSe$_2$ one has to perform two micro-reflectance measurements, one to collect the light reflected from the chosen WSe$_2$ area ($I_{Flake}$) and the other one for the substrate ($I_{Sub}$). The top panel of Figure 2b shows two reflectance spectra recorded at the positions indicated in panel a. The differential reflectance (*D.R.*) of the flake can be calculated according to the formula:

$$D.R. = \frac{I_{Flake} - I_{Sub}}{I_{Flake}}. \qquad (1)$$

In the case of a thin film deposited on a substrate, *D.R.* is proportional to the absorption coefficient of the thin film [19]. The differential reflectance spectrum shown in the bottom panel of Figure 2b displays a peak located at approximately 1.6 eV and various shoulders on a broad increasing background. The peak and the leftmost shoulder have been attributed to the creation of excitons at the K point of the reciprocal space that corresponds to the location of the direct bandgap in TMDCs [11, 20, 21]. The presence of two features separated in energy by few hundreds meV is due to the spin-orbit splitting of the valence band, which is a common feature of 2D TMDCs and it is schematically depicted in the inset of Fig. 2b.

To check the spatial homogeneity of the WSe$_2$ absorption coefficient we record a scanning differential reflectance map. We measure a reflectance spectrum in each point of the area shown in Figure 2a by moving the sample with the motorized x-y stage in steps of 1 µm and then we calculate the differential reflectance spectra according to Equation 1. To extract meaningful information from the map we fit in each spectrum the A exciton peak to a Lorentzian plus a linear polynomial background. Figure 2c shows a map of the A exciton peak energy and Figure 2d a map of the full width at half maximum (FWHM) of the peak. The A exciton energy map of Figure 2c shows that the thinner regions of the flakes have an energy between 1.635 and 1.64 eV (red-yellow color in the map) while the thicker regions have an energy of approximately 1.625 eV (blue). A clear distinction based on the A exciton peak can be made between the different thickness regions. On the other hand, the map of the A exciton FWHM of Figure 2d is more homogeneous between the various regions. In this case mostly green and yellow regions are present corresponding to a FWHM of 60-80 meV, without a clear correlation with the flake thickness. The only region with a different appearance, showing red counts (FWHM larger than 90 meV), is the thin part of the rightmost flake. From the optical image one can see that the flake presents a fold or imperfection there.

To follow the change in the optical properties of single-layer TMDCs induced by the strain we perform differential reflectance measurements as a function of substrate temperature. We first calibrate the thermal expansion of the substrates by taking microscope photographs of the substrate surface at different temperatures after patterning the substrate surface with a periodic structure. By extracting the change in distance between the different structures on the substrate we can plot the change in distance as a function of temperature as shown in Figure 3a. For both PDMS and PP, in the probed temperature range we find a linear increase of the dimensions as a function of temperature. by fitting the data to a linear function we extract the thermal expansion coefficient of each substrate, given by the slope, PDMS = (340 ± 15) $10^{-6}$/K and PP = (136 ± 15) $10^{-6}$/K. The 2D TMDCs have a thermal expansion coefficient of $2 \cdot 10^{-6}$/K [5], approximately two orders of magnitude smaller than the one of PDMS and PP.

Flakes with a single-layer region of $MoS_2$ were deposited onto glass, PDMS and PP substrates. Figure 3b shows a series of differential reflectance spectra of $MoS_2$ recorded at temperatures between 25 °C and 100 °C. Similar to the case already discussed of $WSe_2$ the absorption of $MoS_2$ is dominated by excitons. The spectra on the three substrates and at the different temperatures present similar features. For energies lower than the bandgap the differential reflectance is low, since the absorption is negligible at energies below the bandgap of $MoS_2$. Between 1.85 eV and 2 eV two prominent peaks are present, which are due to the A and B excitons formed in the direct bandgap region of the Brillouin zone. The energy at which these peaks are located depends both on the energy of the bandgap (on the order of 1-2 eV) and on the exciton binding energy (on the order of 10-50 meV). A high absorption tail is present at larger energies due to interband transitions and to the C exciton (not shown in the figure).

From Figure 3b one can see that the center of the A and B peaks shifts at lower energies when increasing the temperature. The rate at which the peaks shift is substrate-dependent, in the case of glass we find A = -0.34 meV/°C and B = -0.42 meV/°C, for PDMS A = -0.42 meV/°C and B = -0.46 meV/°C, while for PP A = -0.70 meV/°C and B = -0.66 meV/°C. The binding energy of the A and B excitons is not expected to change in the experimental tem-perature range, so we attribute the observed redshift to a change in the bandgap energy of MoS$_2$.

The substrate-dependent shift rates are determined by the amount of strain present in MoS$_2$ induced by the substrate added to the intrinsic thermal strain caused by the thermal expansion of MoS$_2$ which is independent from the substrate. When MoS$_2$ is deposited on the glass substrate the shift rate measured is mostly due to the intrinsic MoS$_2$ thermal shift, since the thermal expansion of glass is much lower than the one of MoS$_2$. In the cases of PDMS and PP, one would expect a larger shift of the MoS$_2$ bandgap on PDMS, since its thermal expansion is more than three times larger than the one of PP, while clearly the observed shift rate on PP is much larger. To correct this temperature effect we subtracted the intrinsic thermal expansion of MoS$_2$ from the experimental rates. We find that the strain induced shift in the MoS$_2$ bandgap on PDMS is zero within the experimental error (-0.06 ± 0.04) meV/°C corresponding to a gauge factor of -1.8 meV/%. On the other hand on PP the strain-induced redshift of the bandgap takes the value (-0.3 ± 0.04) meV/°C which translates to -22.1 meV/%.

The large difference between the strain-induced bandgap shifts of MoS$_2$ on PP and PDMS can be explained by the two different substrates Young's moduli, which give different strain transfer efficiencies. To test this hypothesis we perform a finite elements simulation with COMSOL. The inset of Figure 4a shows a cross-section of the model used, which consists of a cylindrical substrate (whose Young's modulus, $E_{Sub}$, is a parameter in the simulation) and a 10 um diameter quasi-2D flake which represents single-layer MoS$_2$ deposited on top. The thickness and lateral dimensions of the substrate are much larger than the top flake. In the simulation we study the strain generated in the top MoS$_2$ layer as a function of the substrate Young's modulus given a fixed biaxial strain of 1% in the substrate. Figure 4a shows the amount of strain transferred to MoS$_2$ in a semi-logarithmic representation. When $E_{Sub}$ is much lower than $E_{MoS2}$, the strain transferred to MoS$_2$ is negligible, like in the case of PDMS. For values $E_{Sub}$ larger than 0.1 GPa the biaxial strain begins to be transferred. A strain transfer close to 100% is achieved at Young's moduli larger than 10 GPa. The Young's modulus of PP (1-2 GPa) is located in the region of medium strain transfer.

Thus PP, thanks to its Young's modulus and large enough thermal expansion coefficient, can effectively transfer biaxial strain to single-layer TMDCs. Figure 4b shows the energy of the A exciton peak of the four single-layer TMDCs as a function of the PP substrate strain. The plot has been compiled by fitting the A peak in the various differential reflectance spectra to extract its position. From the figure we find that the magnitude of the bandgap tuna-bility as a function of biaxial strain follows the trend $WS_2$ > $WSe_2$ > $MoS_2$ > $MoSe_2$, with tungsten and sulfur atoms giving more strain-susceptibility than molybdenum and selenium atoms respectively. According to these findings, the largest tunability, which can be interesting for applications, can be attained using heavier metal atoms and lighter chalcogenide atoms.

Finally we discuss a real-time application of strain, which is not possible with conventional bulk semiconductors. Exploiting the thermal expansion we can easily achieve a 10% modulation of the reflection for certain wavelengths in a time-scale of a few seconds. To demonstrate this we measure the differential reflectance of a $MoS_2$ single-layer flakes deposited on PDMS while cycling the temperature between 40 °C and 50 °C, applying a square wave modulated voltage to the Peltier heater. Figure 5a displays a color-map of the differential reflectance recorded as a function of time. The trace displays reproducibility in time both for what concern the absolute reflectance values and for the position of the A and B excitons. Figure 5b shows the differential reflectance at two wavelengths as a function of time. The signals have a distorted square wave profile with rise/fall characteristic times of 10 seconds with a modulation of the intensity of 8% at 663 nm. The observed modulation in the reflectance is noteworthy especially considering the atomic thickness (< 1 nm) of these single-layer TMDCs. Modulations up to 25% could be reached thanks to the excitons that dominate the dielectric function of single-layer TMDCs materials even at room temperature.

**Conclusion**

In conclusion, we have exploited the large thermal expansion coefficient of polymeric substrates to apply large biaxial tensile strain to single-layer flakes of four transition metal dichalcogenides deposited on top. By recording the differential reflectance spectra of these materials as a function of external strain we monitored the change in bandgap induced by the strain. We observe that the magnitude of the induced energy shift follows the order $MoSe_2$ < $MoS_2$ < $WSe_2$ < $WS_2$ and we demonstrated that a real-time modulation is possible.

**ACKNOWLEDGEMENTS**

AC-G acknowledges funding from the European Commission under the Graphene Flagship, contract CNECTICT-604391. RF acknowledges support from the Netherlands Organisation for Scientific Research (NWO) through the research program Rubicon with project number 680-50-1515. DPdL acknowledges support from the MINECO (program FIS2015-67367-C2-1-P).

**COMPETING INTERESTS**

The authors declare no competing financial interests.

**FUNDING**

A.C.G. European Commission under the Graphene Flagship: contract CNECTICT-604391

R.F. Netherlands Organisation for Scientific Research (NWO): Rubicon 680-50-1515

D.P.dL. MINECO: program FIS2015-67367-C2-1-p

**FIGURES**

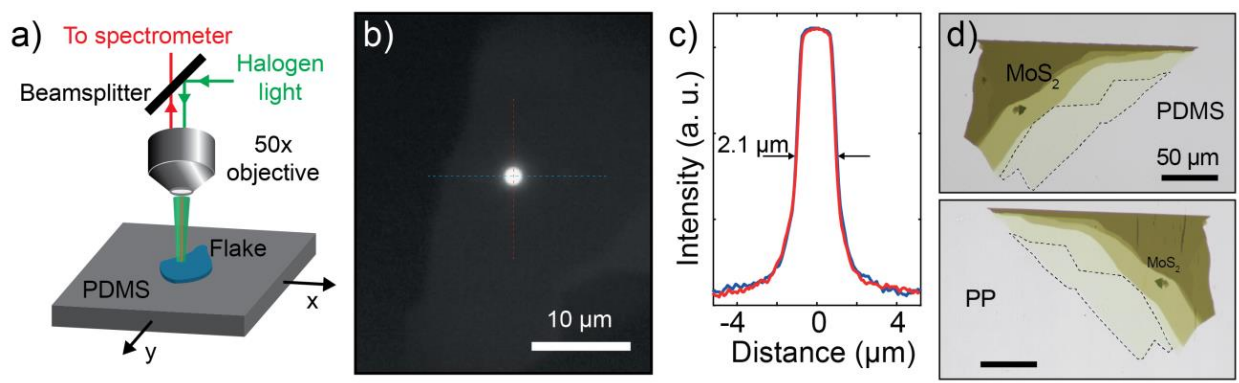

**Figure 1**: a) Schematic illustration of the micro-reflectance setup. b) Photograph of the analyzer spot-size of the setup. c) Line-cuts of the intensity of the spot-size in the horizontal and vertical directions. d) Microscope photographs of a $MoS_2$ flake recorded in transmission illumination mode. The flake initially exfoliated onto a PDMS surface (top) has been deterministically transferred to a PP surface (bottom).

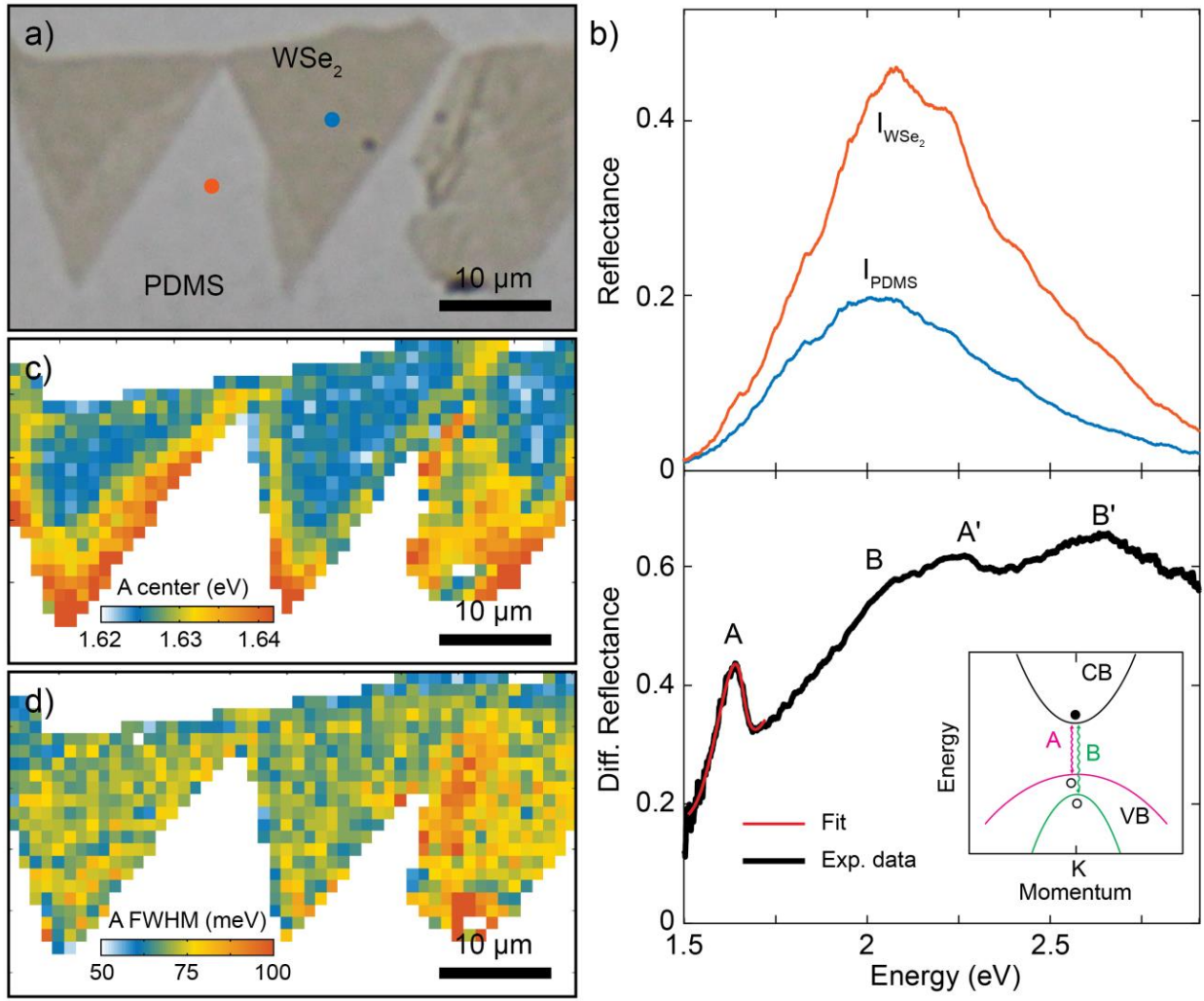

**Figure 2**: a) Microscope photograph of few WSe$_2$ flakes on a PDMS surface. b) Top: micro-reflectance spectra of WSe$_2$ (red) and PDMS (blue) recorded at the positions indicated by the circles in panel (a). Bottom: differential reflectance spectrum of WSe$_2$ calculated from the top two reflectance spectra. Inset: schematic representation of the valence (VB) and conduction (CB) bands at the K point of the reciprocal space of single-layer TMDCs. c) Color-map representing the A exciton peak energy compiled by fitting the A exciton peak in the differential reflectance spectra recorded in each position in a scanning differential experiment. The step-size used for the scanning is 1 μm and each individual spectrum is integrated for 2 seconds. d) Color-map representing the A exciton peak full width at half maximum (FWHM).

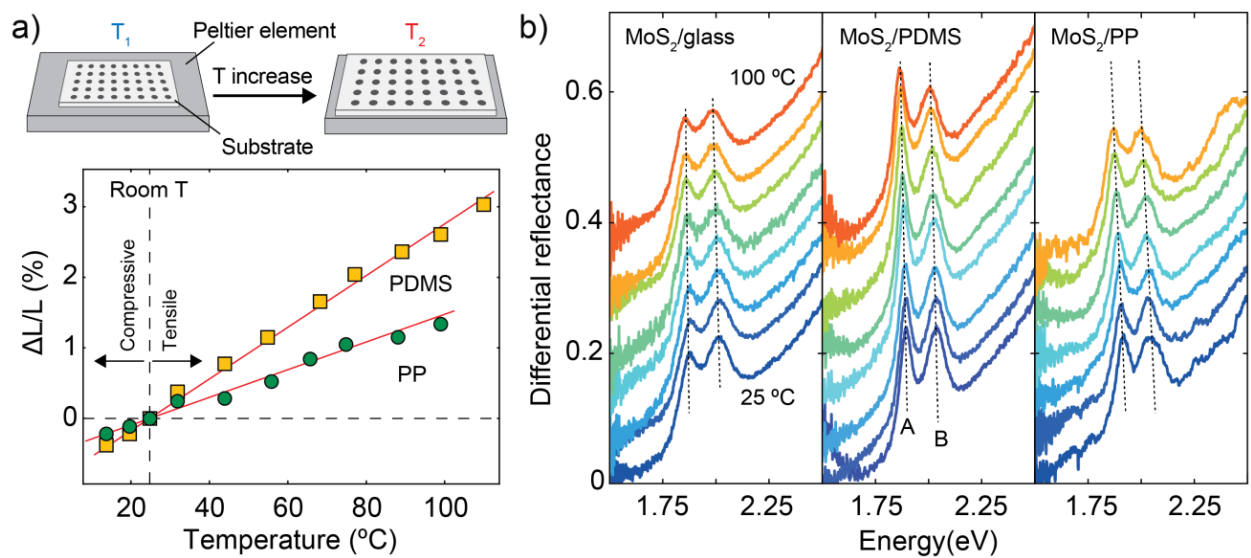

**Figure 3**: a) Top: schematic of the substrate expansion calibration experiment. Bottom: thermal expansion/compression of PDMS and PP substrates. The red lines are fit to the data used to extract the thermal expansion coefficient of each material. b) Differential reflectance spectra of single-layer $MoS_2$ as a function of temperature for three different substrates. The spectra are offset in the vertical direction for clarity.

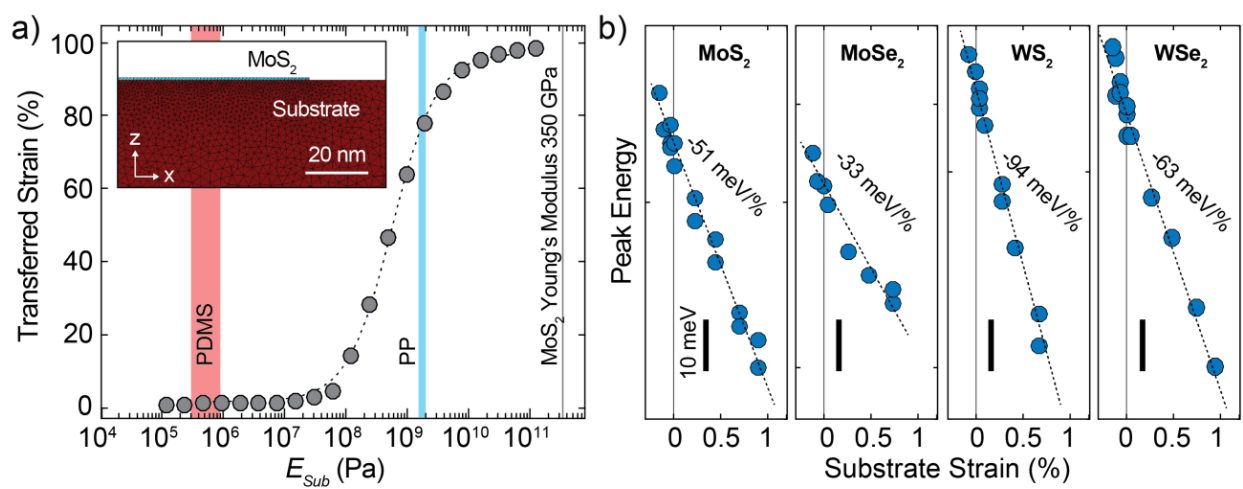

**Figure 4**: a) Maximum transferred strain in $MoS_2$ as a function of substrate's Young's Modulus. Inset: cross-section of the model used consisting of a thick substrate with on top a single-layer $MoS_2$ (0.7 nm thickness) on the substrate with the mesh drawn in black. b) Energy

of the A exciton as a function of substrate strain extracted from the differential reflectance spectra of the four single-layer TMDCs deposited on PP.

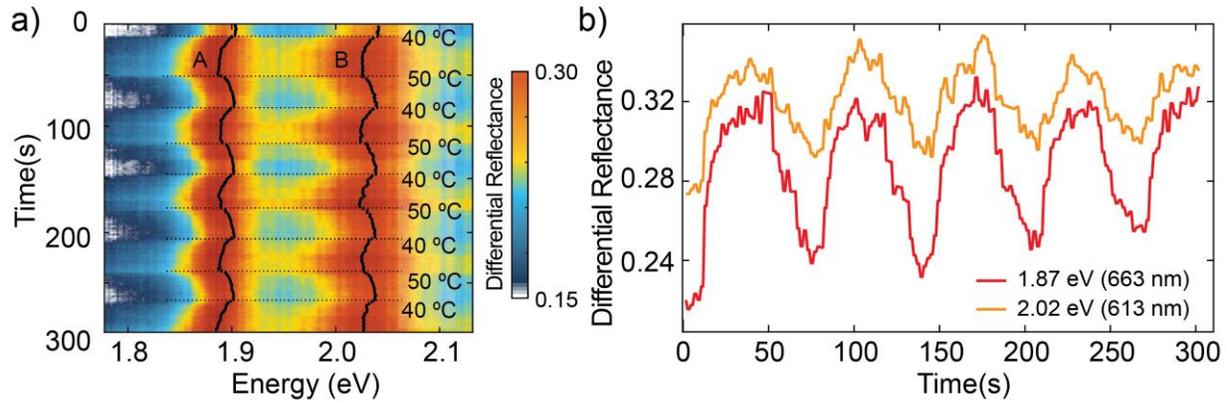

**Figure 5**: a) Color-map representing the differential reflectance of single-layer $MoS_2$ deposited on PDMS as a function of time with periodical heating and cooling cycles of the substrates from 40 °C to 50 °C. Each differential reflectance spectrum has been integrated for 300 ms. b) Differential reflectance extracted at 1.87 and 2.02 eV as a function of time.